\documentclass{aa}
\newcommand{\be}{\begin{equation}}
\newcommand{\ee}{\end{equation}}
\newcommand{\bd}{\begin{displaymath}}
\newcommand{\ed}{\end{displaymath}}
\newcommand{\gapprox}{\;\rlap{\lower 2.5pt                              
             \hbox{$\sim$}}\raise 1.5pt\hbox{$>$}\;}                    
\newcommand{\lapprox}{\;\rlap{\lower 2.5pt                              
             \hbox{$\sim$}}\raise 1.5pt\hbox{$<$}\;}

\newcommand{\mk}{}    

\usepackage{graphicx}
\begin{document}
   \thesaurus{06    
              02.01.2 
              02.08.1 
              08.02.2 
              08.09.2 EX\,Dra
              }
   \title{Spirals and the size of the disk in EX\,Dra}

%
   \author{V. Joergens \inst{1,2} \and  H.C. Spruit \inst{1}
   \and R.G.M. Rutten \inst{3,4}}

 
   \institute{Max-Planck-Institut f\"ur Astrophysik, Postfach 1523, 
               D-85740 Garching bei M\"unchen, Germany \and
               Max-Planck-Institut f\"ur Extraterrestrische Physik, 
               Postfach 1603, D-85740 Garching bei M\"unchen, Germany \and
               Netherlands Foundation for Research in Astronomy \and
               Royal Greenwich Observatory, Apartado de Correos 321, 
               38780 Santa Cruz de La Palma, Canary Islands, Spain
               }

   \date{Received }

   \maketitle



\begin{abstract}
Observations at high spectral and temporal resolution are presented of
the dwarf nova EX\,Dra in outburst. 
The disk seen in the \ion{He}{i} line reconstructed by Doppler tomography
shows a clear two-armed spiral pattern pointing to spiral shocks in the disk.
The Balmer and \ion{He}{ii} maps also give evidence for 
the presence of spirals. 
The eclipse as seen in the red continuum indicates a disk radius of 0.31 times
the orbital separation, which might be
large enough to explain the observed spiral shocks through
excitation by the tidal field of the secondary. 
The eclipse in the Balmer line profiles, well 
resolved in our observations, indicates a somewhat smaller disk size (0.25).
We discuss the possibility that this is related to an optical depth effect 
in the lines.
\end{abstract}

\section{Introduction}

Spiral shocks in accretion disks have been predicted from numerical 
simulations (Sawada et al. 1986, 1987, R\'o\.zyczka \& Spruit 1993, Yukawa et 
al. 1997), and analytic considerations (Spruit 1987, Spruit et al. 1987, 
Larson 1990). They are excited by the tidal field of the secondary if the 
disk extends 
far enough into the Roche lobe and can result in two prominent spiral arms.
If shock dissipation is the main mechanism damping the wave, it extends over 
the entire disk and causes accretion at an effective $\alpha$-value of 
$0.01(H/r)^{3/2}$ (Spruit 1987, Larson 1990, Godon 1997). 

The first observational evidence for shock waves in accretion disks of 
cataclysmic variables (CVs) was the detection of a clear two-armed 
structure in the disk of IP\,Peg during rise to outburst (Steeghs et 
al. 1997). The spiral pattern, interpreted as evidence for shock waves,
has also been seen during outburst maximum (Harlaftis et al. 1999) and early 
decline of outburst (Morales-Rueda et al. 2000).

At the temperatures expected from dwarf nova disks models, whether in outburst 
or in quiescence, the predicted spirals are tightly wound and would be hard to 
detect observationally (Bunk et al. 1990), so that their presence in the 
observations is somewhat unexpected.
Spirals this strong are most naturally explained if the disk
temporarily extends rather far into the primary Roche lobe, so that
the tidal force of the secondary causes a strong disturbance. A strong
non-axisymmetric disturbance, however, would also cause the gas to
loose angular momentum quickly (transfered to the secondary), so that
the disk would shrink to a smaller radius where the tidal force is
weaker. Spirals in disks of CVs would then be understandable if they are a
temporary phenomenon, perhaps restricted to
outbursts. To test this, more observations of different systems at
sufficient spectral resolution and signal-to-noise are needed
(Steeghs \& Stehle 1999). 

With high quality spectroscopic studies of different CVs it should also be  
possible to answer the question if spiral shocks in CV accretion disks are a 
common phenomenon.  Systems suitable for this purpose would be bright and 
have frequent outbursts, such as SS\,Cyg and EX\,Dra. 

EX\,Dra is a double--eclipsing dwarf nova with a 5-hour orbit 
(Barwig et al. 1993, Billington et al. 1996, system parameters by 
Fiedler et al. 1997).
There is suggestive evidence for asymmetric structures in the \ion{He}{i}
Doppler map reconstructed from outburst data taken in 1993  (Joergens et al. 
2000). From eclipse maps obtained at various stages in the outburst cycle Baptista \& 
Catalan (1999) claim that spiral waves form at the early stages of an 
outburst. We report in this paper on observations at high spectral and temporal
resolution during an outburst in 1996.

\section{Observations and reduction}
EX\,Dra was observed on the nights of July 27
and 28 1996, with the ISIS spectrograph on the 4.2\,m William
Herschel Telescope. 
The red ISIS arm was equipped with the TEK\,5 CCD,
the blue arm with TEK\,1 CCD, covering the 
wavelength ranges 6375-6778\,{\AA} and
4585-4993\,{\AA} respectively, at a dispersion of
0.4\,{\AA}/pixel. 
The spectra were observed during seeing of 1.4 -- 1.7 arc sec, 
using an exposure time of 60s.
The spectra were optimally-extracted, including the elimination of 
cosmic ray hits but not the correction of Pixel-to-pixel variations of the 
detector sensitivity. 
Slit losses due to variable atmospheric conditions were 
corrected using a faint comparison star on the slit.
The red spectra, but not the blue spectra, were flux calibrated. 

Our spectra were taken roughly in the middle of an outburst. Photometric 
observation at the Wendelstein observatory showed that the system was already 
in outburst on 26 July, and in quiescence on 4 August (Barwig, private comm.). 
VSNET (1998) records show that the system was in quiescence on 24 July 
and in outburst one day later and in quiescence again on 1 August. The outburst 
therefore started on 25 August, three days before our observations.

\section{Interpretation of the Doppler maps}
\label{interpret}

Doppler maps were computed from the phase-folded spectra of 28 July, using the IDL-based 
fast-maximum entropy package described by Spruit (1998)\footnote{available 
at http:www.mpa-garching.mpg.de/$\sim$henk}. 
For further details on Doppler tomography see Marsh \& Horne (1988). 
{\mk The eclipsed part of the data (phases -0.12 to 0.12) has been 
excluded from the reconstruction process, and does not affect the maps produced.
For information, however, these parts of the data are included in the spectra
shown in Fig.~1. 
They were used separately to derive disk sizes in the lines and
the continuum.}

\begin{figure*}
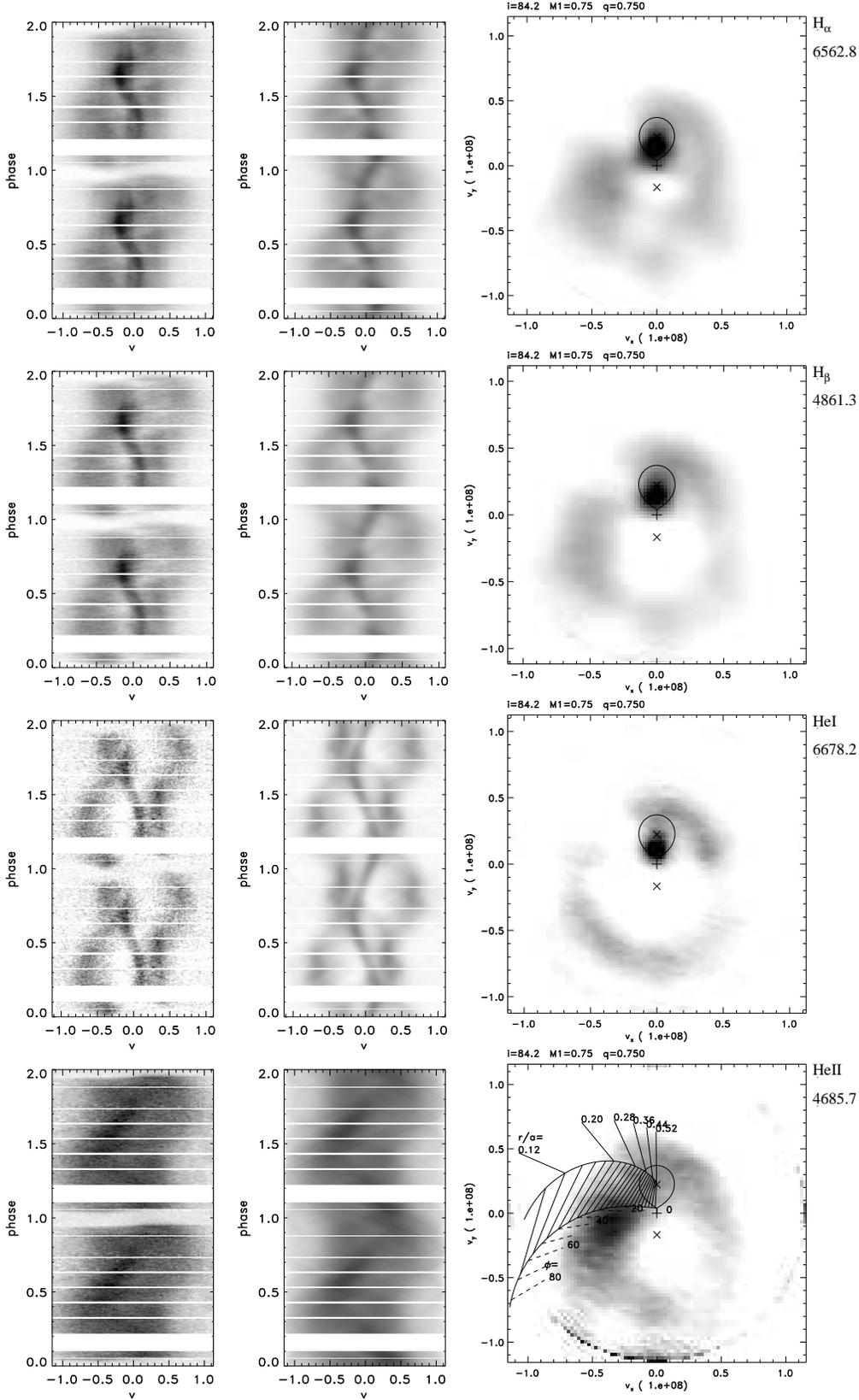

\vbox{
\hbox{\mbox{}\hfill
 \includegraphics[width=5.3cm,height=3.5cm,angle=90]{bh033.f1}
 \hfill\includegraphics[width=5.3cm,height=3.5cm,angle=90]{bh033.f2}
 \hfill\includegraphics[height=5.4cm]{bh033.f3}\hfill\mbox{}}
\hbox{\mbox{}\hfill\includegraphics[width=5.3cm,height=3.5cm,angle=90]
{bh033.f4}
 \hfill\includegraphics[width=5.3cm,height=3.5cm,angle=90]{bh033.f5}
 \hfill\includegraphics[height=5.4cm]{bh033.f6}\hfill\mbox{}}
\hbox{\mbox{}\hfill\includegraphics[width=5.3cm,height=3.5cm,angle=90]
{bh033.f7}
 \hfill\includegraphics[width=5.3cm,height=3.5cm,angle=90]{bh033.f8}
 \hfill\includegraphics[height=5.4cm]{bh033.f9}\hfill\mbox{}}
\hbox{\mbox{}\hfill\includegraphics[width=5.3cm,height=3.5cm,angle=90]
{bh033.f10}
 \hfill\includegraphics[width=5.3cm,height=3.5cm,angle=90]{bh033.f11}
 \hfill\includegraphics[height=5.4cm]{bh033.f12}\hfill\mbox{}}
}
\caption{\label{colla} Phase-folded spectra of EX\,Dra in outburst on 28 July 
1996 
(left column), corresponding Doppler maps (right column), and the spectra 
reconstructed from the maps (middle). Top to bottom: H$_\alpha$, H$_\beta$,
\ion{He}{i}, \ion{He}{ii}. The mass transferring stream (see text) and
the Doppler 
image of the Roche lobe of the secondary star are over plotted on the 
\ion{He}{ii} 
image. Upper and lower intensity cuts of the Doppler maps have been adjusted 
such that the disk emission has roughly the same contrast in each image.
}
\end{figure*}

Fig. \ref{colla} shows the results for the four strongest lines in the spectra,
H$_\alpha$, H$_\beta$, \ion{He}{i}\,$\lambda 6678$ and \ion{He}{ii}\,$\lambda 
4686$.  The phase-folded spectra are shown in the left column, the 
corresponding Doppler maps in the right and in the middle the spectra 
reconstructed from the maps. The theoretical
trajectory of the mass transfering stream
has been plotted in the \ion{He}{ii} image, together with the Keplerian 
velocity along the stream path (cp. Marsh\,\&\,Horne 1988). Bars
connecting the two arcs indicate correspondence in physical space, and are 
annotated with radius r/a and the azimuth $\Phi$ relative to the primary.
The system parameters  
and the ephemeris are from Fiedler et al. (1997).

The \ion{He}{i} image shows the spirals clearest:
asymmetric disk emission is concentrated 
in two arms in the first and third quadrant. 
We interpret this deviation from a Keplerian disk
as indication of the presence of spiral shock waves in the accretion 
disk. The image looks quite similar to
the \ion{He}{i} map of IP\,Peg published by Steeghs et al. (1997). 
As in the case of IP\,Peg, the arms do not follow a circle centered on the white 
dwarf, but are somewhat elongated along the V$_y$-axis. 
This is the pattern expected from spiral shocks (Steeghs 
\& Stehle 1999). The asymmetry of the spirals in the EX\,Dra disk  
is smaller than in the IP\,Peg observations
(Steeghs et al. 1997, Harlaftis et al. 1999, Morales-Rueda et al. 2000), 
perhaps indicating that the shocks are weaker.
The spiral arm in the upper right is stronger in intensity as well as in
asymmetry than that one in the lower left. This pattern is also seen in the 
Doppler maps of IP\,Peg. 
As in the case of IP\,Peg, the Balmer lines show similar, 
but less clearly defined structures.

The Doppler images, with the exception of \ion{He}{ii},  
show strong emission from the secondary star, also visible during the 
outburst in 1993 (Joergens et al. 2000). This indicates 
heating of the secondary by radiation from the inner disk. 
The temperatures are obviously 
not high enough to excite the \ion{He}{ii} line. 
The \ion{He}{ii} line also differs 
from the other lines by the presence of a prominent emission patch 
at the theoretical gas stream trajectory.
The center of this hot spot emission is somewhat below the trajectory,
consistent with quiescence observations of EX\,Dra (Billington et al.
1996, Joergens et al. 2000). 

\section{The disk radius}
\label{disk}
The large width of the \ion{He}{ii} line is as expected for a high
excitation line produced near the center of a disk in outburst. The
eclipse of the \ion{He}{ii} line is visible in the spectrum up to a velocity
of about 1500\,km/s. The orbital velocity at the disk edge is about 4.3
times lower than this. Assuming Keplerian rotation, the disk region
eclipsed at 1500\,km/s 
is about 20 times smaller in radius than the outer radius 
of the disk. At our phase resolution, the 1500 km/s emission is thus 
point-like, 
and can be used to measure the width of the secondary's shadow on the orbital 
plane, $\Delta\phi=0.08\pm0.005$. From the position of the hot spot in the \ion{He}{ii}
map, we can estimate a disk size of $r/a=0.32 \pm 0.04$. 

The eclipse of the Balmer and \ion{He}{i} lines shows that the blue
disk emission reappears (at V$\approx 350$km/s) at the same phase as
the last red emission disappears, within 
the measurement uncertainty. The size of the disk as seen in these lines is thus,
by a coincidence, nearly as wide as the occulting shadow of the secondary. 
Hence the disk radius as seen in the Balmer lines is 
$r_{\rm dB}/a\approx 2\pi\Delta\phi/2=0.25\pm0.02$. 

We can compare this with the disk size as measured from the eclipse in the
continuum. The light curve in the continuum between 6400 and 6500{\AA}, 
extracted from our spectra, is shown in Fig. 2. The width of the eclipse is 
$w=0.17\pm 0.005$ orbits. With $\Delta\phi=0.08$, this implies a disk radius 
$r_{\rm d}/a=0.31\pm 0.01$.

Fiedler et al. (1997) find a value of 0.11 for the eclipse width of
the white dwarf, from analysis of a mean blue light curve in quiescence. The 
disk as seen in the continuum thus appears to be larger in outburst than in 
quiescence, as expected. 
The eclipse width values of 0.17 in outburst and 0.11 in 
quiescence are also compatible with the eclipse light curves of Baptista \& 
Catalan (1999). These authors measured the eclipses of EX\,Dra
at various stages in the 
outburst cycle.

\begin{figure}
 \includegraphics[height=0.49\textwidth,angle=270]
{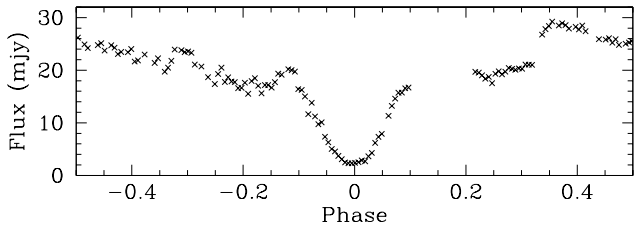}
\caption{Spectrophotometric light curve of the wavelength range 
6400--6500{\AA}.}
\end{figure}

\section{Discussion and conclusions}
We find evidence for spiral structures in the outburst accretion disk
of EX\,Dra similar to those found in IP\,Peg by Steeghs\,et\,al. (1997). 
The pattern of intensity and velocity perturbations agrees with that predicted 
from numerical simulations of spiral shock waves (Steeghs\,\&\,Stehle 1999). It
is best seen in the \ion{He}{i} line, somewhat less clearly in the Balmer and 
\ion{He}{ii} lines. 
In EX Dra the pattern appears somewhat less clearly and
asymmetric than in IP\,Peg.

In particular, it is less clear in the 
\ion{He}{ii} line, suggesting lower temperatures and shock strengths
of the spirals of EX\,Dra. 
Possibly the observability of spirals in the \ion{He}{ii} map 
is hampered by strong hot spot emission visible in this line.

We derive a disk radius 
of $r/a=0.31\pm0.01$ from the  red continuum eclipse light curve.
From numerical simulations, Steeghs \&
Stehle find that disk sizes \mbox{$r_{\rm d}/a\,=\,0.3-0.4$} 
are needed to excite 
spirals that are strong enough to generate an observable pattern in the 
spectra.
(Transformation from $r_{\rm d}/a$ to $r_{\rm d}/R_{L_1}$ is given by
R$_{L_1}=0.53\,a$ for q=0.75, cp. Plavec \& Kratochvil 1964.)
Since the tidal force is a very steep function of $r/a$, the 
spirals rapidly become weak at smaller disk sizes. The {\mk disk 
size we find here in EX\,Dra} is at the lower limit of the required size. 

Further evidence for the size of the disk in EX\,Dra in outburst was
obtained by Baptista \& Catalan (1999). 
Radial intensity distributions presented there
show disk radii of 0.30\,$a$ in quiescence
and 0.49\,$a$ in outburst. The authors see hints of spirals in their 
eclipse maps, during the early outburst stages.
This may be compared with the 
observations presented here, which show spirals and a 
disk size of 0.31\,$a$ three days after the beginning of an outburst.
 
The eclipses of the Balmer lines in our spectra yield significantly {\em smaller} 
disks sizes than the continuum eclipse, $r_{\rm dB}/a=0.25\pm 0.02$.
Since it is known that the line emission is produced primarily in the outer
parts of the disk (e.g. Rutten et al. 1994), one might have expected the disk 
as seen in the lines to be larger, if anything, than in the continuum. 

A possible resolution of this conflict may lie in the optical depth effects 
affecting the emission lines in systems seen at high inclination. {\mk The low central 
intensity of the Balmer lines in high-inclination CVs (often below the continuum)
shows that such effects are strong. As shown by Horne \& Marsh (1986),
the effects are strongest for lines of sight parallel and perpendicular
to the orbital motion, leading to reduced line emission from these directions
compared to intermediate lines of sight (near $45^\circ$ to the orbit).
The bias towards intermediate angles will give the appearance of a somewhat
smaller disk size. It is still to be determined if this suggestion also
works quantitatively.}


\begin{acknowledgements}
We thank Dr.\ Heinz Barwig for providing the photometric information on
EX\,Dra. We thank the anonymous referee 
for the comments, which helped to improve the presentation of the data and to 
correct an error. This work was done in the context of the research network 
`Accretion onto black holes, compact objects and protostars' (TMR Grant 
ERB-FMRX-CT98-0195). The Isaac Newton 
Telescope is operated on the island of La Palma by the Isaac Newton Group of 
Telescopes in the Spanish Observatorio del Roque de los Muchachos of the 
Instituto de Astrofisica de Canarias.
\end{acknowledgements}

\end{document}